\documentclass{ws-procs11x85}

\usepackage{balance}

\newcommand*{\bz}{\ensuremath{{B^0}}}

\newcommand*{\piz}{\ensuremath{{\pi^0}}}
\newcommand*{\pip}{\ensuremath{{\pi^+}}}
\newcommand*{\pim}{\ensuremath{{\pi^-}}}
\newcommand*{\kp}{\ensuremath{{K^+}}}
\newcommand*{\km}{\ensuremath{{K^-}}}
\newcommand*{\ks}{\ensuremath{{K_S^0}}}
\newcommand*{\kl}{\ensuremath{{K_L^0}}}
\newcommand*{\kstarz}{\ensuremath{{K^{*0}}}}
\newcommand*{\jpsi}{\ensuremath{{J/\psi}}}

\newcommand*{\fcp}{\ensuremath{{f_{CP}}}}

\newcommand*{\Nev}{\ensuremath{{N_\textrm{ev}}}}

\newcommand*{\taubz}{\ensuremath{{\tau_\bz}}}

\newcommand*{\fq}{\ensuremath{q}}

\newcommand{\cala}{{\cal A}}
\newcommand{\cals}{{\cal S}}

\makeindex
\begin{document}

\title{Results on the CKM angle {\boldmath $\phi_1$ $(\beta)$}}

\author{T.~E. Browder}

\address{Department of Physics and Astronomy, University of Hawaii,
2505 Correa Road, Honolulu, HI 96822, USA\\E-mail: teb@phys.hawaii.edu}

%%%%%%%%%%%%%%%%%%%%%%%%%%%%%%%%%%%%%%%%%%%%%%%%%%%%%%%%%%%%%%%%%%%%%%%%%
% You may repeat \author \address as often as necessary             %
%%%%%%%%%%%%%%%%%%%%%%%%%%%%%%%%%%%%%%%%%%%%%%%%%%%%%%%%%%%%%%%%%%%%%%%%%

\twocolumn[\maketitle\abstract{I review results related
to the CKM angle $\phi_1$($\beta$).
These results include recent measurements of $CP$-violation from the BaBar
and Belle experiments in $b\to c \bar{c} s$, $b\to c \bar{c} d$ and
$b\to s q \bar{q}$ processes.}]

\baselineskip=13.07pt
\section{Introduction}
\subsection{The B Physics Program}\label{subsec:program}

The $B$ physics program addresses several fundamental
 questions. Is the irreducible
phase in the Cabibbo-Kobayashi-Maskawa (CKM) matrix the source of
all $CP$-violating phenomena in the $B$ system?\cite{bib:KM} Or is $CP$-violation,
the first manifestation of physics beyond the Standard Model? A related
question is whether there are new $CP$-violating phases from physics
beyond the Standard Model\rlap.\,\cite{grossman}

The unitarity of the CKM matrix implies the existence of three
measurable phases. In the convention favored at KEK and Belle, 
these are denoted
\begin{equation}
\phi_1\equiv arg
\left( \begin{array}{c}
-\frac{V_{cd}V^*_{cb}}{V_{td}V^*_{tb}}
\end{array} \right)
\end{equation}
\begin{equation}
\phi_2\equiv arg
\left( \begin{array}{c}
-\frac{V_{ud}V^*_{ub}}{V_{td}V^*_{tb}}
\end{array} \right)
\end{equation}
\begin{equation}
\phi_3\equiv arg
\left( \begin{array}{c}
-\frac{V_{cd}V^*_{cb}}{V_{ud}V^*_{ub}}
\end{array} \right).
\end{equation}
while at SLAC and at BaBar these angles are usually
referred to as $\beta, \alpha$ and $\gamma$, respectively.

As first noted by Bigi, Carter and Sanda\rlap,\,\cite{bib:sanda}
 there are large measurable $CP$-asymmetries
in the decays of neutral $B$ mesons to $CP$-eigenstates.
In the decay chain $\Upsilon(4S)\to B^0 \bar{B}^0 \to f_{CP}f_{\rm tag}$,
where one of the $B$ mesons decays at time $t_{CP}$ to a 
final state $f_{CP}$ 
and the other decays at time $t_{\rm tag}$ to a final state  
$f_{\rm tag}$ that distinguishes between $B^0$ and $\bar{B}^0$, 
the decay rate has a time dependence
given by\cite{bib:sanda}
\begin{eqnarray*}
\label{eq:psig}
% {\cal P}(\Delta{t})  = & \\
 \frac{e^{-\frac{|\Delta{t}|}{\taubz}}}{4{\taubz}} 
 \biggl\{1 + \fq\cdot 
\Bigl[ \cals\sin(\Delta m_d\Delta{t})
   + \cala\cos(\Delta m_d\Delta{t})
\Bigr]
\biggr\}, 
\end{eqnarray*}
where $\tau_{B^0}$ is the $B^0$ lifetime, $\Delta m_d$ is the mass difference 
between the two $B^0$ mass
eigenstates, $\Delta{t}$ = $t_{CP}$ $-$ $t_{\rm tag}$, and
the $b$-flavor charge $\fq$ = +1 ($-1$) when the tagging $B$ meson
is a $B^0$ ($\bar{B}^0$).
The $CP$-violation parameters $\cals$ and $\cala$  are given by
\begin{equation}
\cals \equiv \frac{2{\cal I}m(\lambda)}{|\lambda|^2+1}, \qquad
\cala \equiv \frac{|\lambda|^2-1}{|\lambda|^2+1},
\end{equation}
where $\lambda$ is a complex 
parameter that depends on both the $B^0 \bar{B}^0$
mixing and on the amplitudes for $B^0$ and $\bar{B}^0$ to decay to
$f_{CP}$. To a good approximation,
the SM predicts $\cals = -\xi_f\sin 2\phi_1$,  where $\xi_f = +1
(-1)$ corresponds to  $CP$-even (-odd) final states.
Direct $CP$-violation,  ${\cal A} =0$ (or 
equivalently $|\lambda| = 1$), is expected 
 for both $b \to c\overline{c}s$ and $b \to s\overline{s}s$ transitions.

\subsection{Accelerators and Detectors}\label{subsec:accdetect}

The $B$-factory accelerators, PEPII\cite{bib:PEPII} and KEKB\cite{bib:KEKB}
were commissioned with remarkable speed starting in late 1998. The
experiments, BaBar\cite{bib:BaBar} 
and Belle\rlap,\,\cite{bib:Belle}  started physics data taking in 1999.
In the summer of 2001, the two experiments
announced the observation of the
first statistically significant signals for $CP$-violation
outside of the kaon system\rlap.\,\cite{bib:belle_prl,bib:babar_prl}

Due to the extraordinary performance of the two accelerators,
the most recent results reported in the summer of 2003 
at the Lepton-Photon Symposium are based on very
large data samples. BaBar has integrated 113 fb$^{-1}$ on the 
$\Upsilon(4S)$ resonance while Belle has integrated a sample of 140
fb$^{-1}$. 
KEK-B also passed a critical milestone for $e^+ e^-$ 
storage rings and 
achieved a peak luminosity above $1\times 10^{34}$~cm$^{-2}s^{-1}$.

\subsection{The Principle of the Measurement}

The measurement of time-dependent $CP$-asymmetry
requires:
\begin{itemize}
\item A large sample of $\Upsilon(4S)$ decays into $B^0\bar{B}^0$ pairs.
To boost the $\Upsilon(4S)$ decay frame 
so that the $B$ mesons' flight length 
can be measured with solid-state vertex detector technology,
both the KEKB and PEP-II accelerators use asymmetric energy beams
with energies of 8.0 and 3.5 GeV or 9.0 and 3.1 GeV, respectively.
\item Efficient reconstruction of  $B\to X_{c\overline{c}}K^0$
decays. This implies accurate measurements of momenta and energies
of neutrals using CsI(Tl) crystal calorimeters in
addition to good charged particle tracking in small cell
drift chambers and
efficient identification of leptons and $K_S^0$ as well as $K_L^0$ mesons.
\item A measurement of $\Delta t$.
This is related
to the measurement of $\Delta z$, the spatial distance between
the decay vertices and achieved at both experiments by using
double-sided silicon strip detectors situated at small radii close
to the interaction point.
\item
A determination of the flavor of the accompanying $B$ (``tagging'');
this is based on the identification of
electrons, muons and charged kaons and the measurement of their
charge.
\end{itemize}

More detailed descriptions of the detectors\cite{bib:BaBar,bib:Belle}
and the experimental analysis procedure
are available elsewhere\rlap.\,\cite{bib:browder_faccini}

\begin{figure}[t]
\center
\psfig{figure=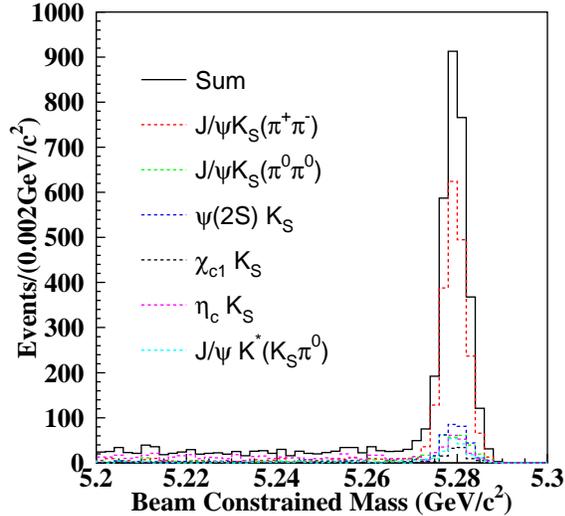,width=8.0truecm}
\caption{The fully reconstructed
$CP$-eigenstate sample used by Belle. This sample
is obtained from a data sample with an integrated luminosity of
140 fb$^{-1}$.}
\label{fig:cpmbc2003}
\end{figure}

\section{Status of $CP$-Violation in $b\to c \bar{c} s$ Processes}

Belle and BaBar reconstruct
 $B^0$ decays to the following $b\to c\bar{c} s$ ${CP}$-eigenstates:
$J/\psi K_S$, $\psi(2S)K_S$, $\chi_{c1}K_S$, $\eta_c K_S$ for
 $\xi_f=-1$  and $J/\psi K_L$ for $\xi_f=+1$\rlap.\,\cite{bib:CC} 
The two classes ($\xi_f=\pm 1$)
should have $CP$-asymmetries that are opposite in sign.

\begin{figure}[t]
\center
\psfig{figure=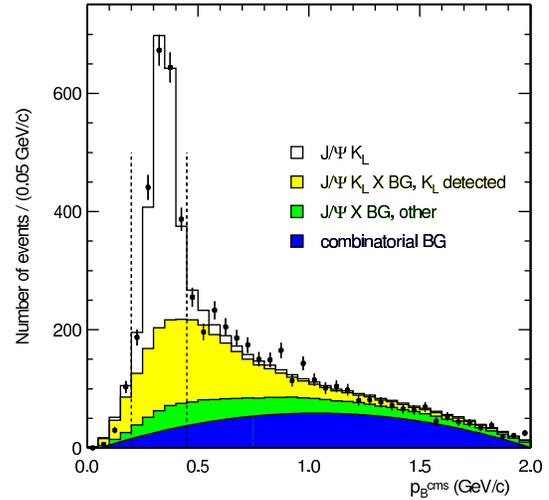,width=7.0truecm}
\caption{The $p_B^*$ ($B$ momentum in the CM frame) distribution for the
$B\to J/\psi K_L$ sample used by Belle. This sample
is obtained from a data sample with an integrated luminosity of
140 fb$^{-1}$. The shaded portions show the contributions
of different background components. The vertical dashed lines
indicate the signal region.}
\label{fig:psikl2003}
\end{figure}

Both experiments also use $B^0\to J/\psi K^{*0}$ decays where
$K^{*0}\to  K_S\pi^0$.
Here the final state is a mixture of even and odd $CP$.
The $CP$ content can, however, be determined from an
angular analysis of other $\psi K^*$ decays. The $CP$-odd
fraction is found to be small (i.e. ($19\pm 4$)\% (($16\pm 3.5$)\%)
in the Belle (BaBar) analysis).

The most recent
BaBar analysis is based on a data sample with an integrated
luminosity of 81 fb$^{-1}$ and 
was first presented in 2002\rlap.\,\cite{bib:babar_prl} 
There is a corresponding published Belle result also shown in 2002
with 78 fb$^{-1}$\rlap.\,\cite{bib:belle_prl} At this Symposium,
Belle provided a new preliminary result for their
140 fb$^{-1}$ sample\rlap.\,\cite{bib:belle_summer_phi1} 

The data sample used for the recent Belle measurement is
shown in Fig.~\ref{fig:cpmbc2003} and Fig.~\ref{fig:psikl2003}. 
Table~\ref{tab:number} lists the numbers of candidates, $\Nev$,
and the estimated signal purity for each $\fcp$ mode.
It is clear that the
$CP$-eigenstate samples that are used for the $CP$-violation
measurements in $b\to c \bar{c} s$ are large and clean.

%%%%%%%%%%%%%%%%%%%%%%%%
\begin{table}
  \caption{\label{tab:number} The yields from Belle for 
    reconstructed $B \to \fcp$
    candidates after flavor tagging and vertex reconstruction, $\Nev$,
    and the estimated signal purity, $p$, in the signal region for each $\fcp$ mode.
    $\jpsi$ mesons are reconstructed in $\jpsi \to \mu^+\mu^-$ or $e^+e^-$
    decays. Candidate $\ks$ mesons are reconstructed in $\ks \to \pi^+\pi^-$
    decays unless otherwise written explicitly.}
%  \begin{ruledtabular}
    \begin{tabular}{llrl}
\hline
      \multicolumn{1}{c}{Mode} & $\xi_f$ & $\Nev$ & \multicolumn{1}{c}{$p$} \\
      \hline 
      $J/\psi \ks $                & $-1$ & 1997 & $0.976\pm 0.001$ \\
      $J/\psi \ks(\piz\piz)$       & $-1$ &  288 & $0.82~\pm 0.02$ \\
      $\psi(2S)(\ell^+\ell^-)\ks$  & $-1$ &  145 & $0.93~\pm 0.01$ \\
      $\psi(2S)(\jpsi\pip\pim)\ks$ & $-1$ &  163 & $0.88~\pm 0.01$ \\
      $\chi_{c1}(\jpsi\gamma)\ks$  & $-1$ &  101 & $0.92~\pm 0.01$ \\
      $\eta_c(\ks\km\pip)\ks$      & $-1$ &  123 & $0.72~\pm 0.03$ \\
      $\eta_c(\kp\km\piz)\ks$      & $-1$ &   74 & $0.70~\pm 0.04$ \\
      $\eta_c(p\overline{p})\ks$   & $-1$ &   20 & $0.91~\pm 0.02$ \\
      \cline{3-4}
      All with $\xi_f = -1$        & $-1$ & 2911 & $0.933\pm 0.002$ \\
      \hline
      $J/\psi\kstarz(\ks\piz)$ & +1(81\%)
                                          &  174 & $0.93~\pm 0.01$ \\
      \hline
      $J/\psi\kl$                  & $+1$ & 2332 & $0.60~\pm 0.03$ \\
\hline
    \end{tabular}
%  \end{ruledtabular}
\end{table}
%%%%%%%%%%%%%%%%%%%%%%%%
%

In the summer of 2001, the first statistically significant measurements
of the $CP$-violating parameter $\sin 2 \phi_1$ were reported
by Belle and BaBar. Belle found
\begin{equation}
 \sin 2 \phi_1 = 0.99 \pm 0.14 \pm 0.06
\end{equation} 
while BaBar obtained
\begin{equation}
\sin 2 \phi_1 = 0.59 \pm 0.14 \pm 0.05.
\end{equation} 
The results
were based on data samples of comparable size (31 million and 32
million $B \bar{B}$ pairs, respectively).

\begin{figure}[t]
\center
%\rule{2cm}{0.2mm}\hfill \rule{2cm}{0.2mm}
%\vskip 6cm
%\rule{2cm}{0.2mm}\hfill \rule{2cm}{0.2mm} 
\psfig{figure=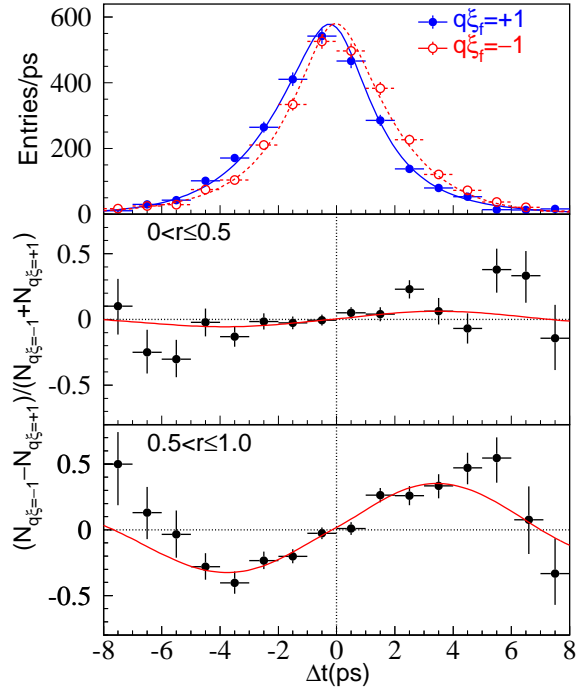,width=8.0truecm}
\caption{Belle data from 2003: (a) $\Delta t$ distributions for $B^0$
and $\bar{B}^0$ tags (b) raw asymmetry
for low-quality tags and (c) raw asymmetry for high-quality tags.
The smooth curves are projections of the unbinned likelihood fit.}
\label{fig:asym_belle}
\end{figure}

The new
Belle data are shown in Fig.~\ref{fig:asym_belle}. This figure shows
the $\Delta t$ distributions where a clear shift between
$B^0$ and $\bar{B}^0$ tags is visible as well as the raw asymmetry plots
in two bins of the flavor tagging quality variable $r$. For low-quality
tags ($0<r<0.5$), which have a large background dilution,
 only a modest asymmetry is visble while in 
the sub-sample 
with high quality tags ($0.5<r<1.0$), a very clear asymmetry with
a sine-like time modulation is present. The final results are extracted
from an unbinned maximum-likelihood fit to the $\Delta t$
distributions that takes into account resolution, mistagging and
background dilution.
The new Belle result with 140 fb$^{-1}$ (152 million $B\bar{B}$
pairs) is 
\begin{equation}
 \sin 2 \phi_1 = 0.733 \pm 0.057 \pm 0.028.
\end{equation} 

The new Belle result may be compared to the BaBar result with 78 fb$^{-1}$
of
\begin{equation}
 \sin 2 \phi_1 = 0.741 \pm 0.067 \pm 0.03 .
\end{equation} 
%The BaBar time distribution data are shown in Fig.~\ref{fig:asym_babar}.
%This figure demonstrates that the $CP$-odd and $CP$-even subsamples have the
%opposite asymmetries, as expected.

%\begin{figure}
%\center
%\psfig{figure=babar_psiks_2002.eps,width=6.0truecm}
%\caption{BaBar data: (a) $\Delta t$ distributions (b) Raw Asymmetry
%for CP odd final states and (c) and (d) show the same distributions
%for CP even final states.}
%\label{fig:asym_babar}
%\end{figure}

\begin{figure}
\center
\psfig{figure=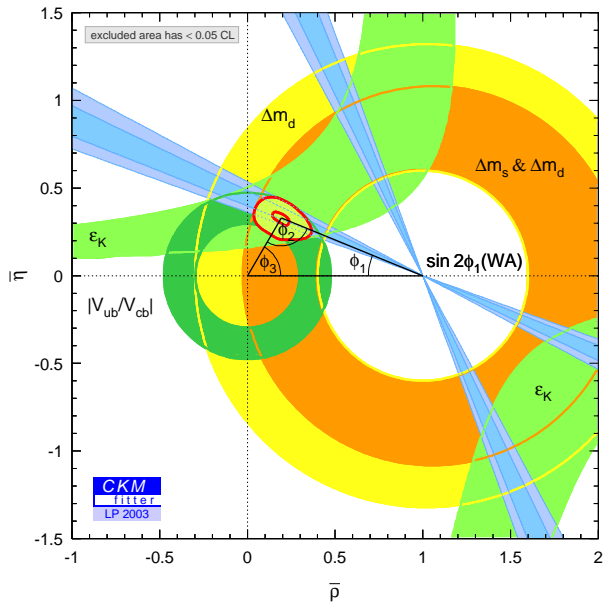,width=8.0truecm}
\caption{Indirect constraints on the angles of the CKM triangle 
compared to the  most recent direct measurements of $\phi_1$ from
Belle and BaBar. The theoretical uncertainties in the indirect
constraints are conservatively estimated by the CKM fitter group.}
\label{fig:ckmfitter}
\end{figure}

Both experiments are now in very good agreement.
A new world average can be calculated from these results,
\begin{equation}
 \sin 2 \phi_1 = 0.736 \pm 0.049.
\end{equation} 
This world average can be interpreted as a constraint
on the CKM angle $\phi_1$. This constraint can be compared
to the indirect determinations on the unitarity triangle\rlap.\,\cite{bib:ckm_fitter}
This comparison is shown in Fig.~\ref{fig:ckmfitter} and 
is consistent with the hypothesis that the Kobayashi-Maskawa
phase is the source of $CP$-violation.

The measurement of $\sin(2\phi_1)$ in $b\to c\bar{c} s$ modes,
although still statistically limited, is becoming a precision
measurement. The systematics are small and well-understood.
Recently, BaBar physicists discovered a new small source of
systematic uncertainty due to $CP$-violation in $b\to c\bar{u} d$
decays on the tagging side\rlap.\,\cite{bib:fbtginterference} 

The presence of an asymmetry with a cosine dependence ($|\lambda| \neq 1$)
would indicate direct $CP$-violation.
In order to test for this possibility in $b\to c\bar{c} s$ modes,
Belle also performed a fit with
$a_{CP} \equiv -\xi_f {\rm Im}\lambda/|\lambda|$
and $|\lambda|$ as free parameters, keeping everything else the
same. They obtain
\begin{eqnarray}
|\lambda| = 1.007\pm 0.041({\rm stat}) \\
a_{CP} = 0.733\pm 0.057{\rm (stat)}, \nonumber
\end{eqnarray} 
for all the $b\to c\bar{c} s$ $CP$ modes combined.
This result is consistent with the assumption used in their
primary analysis.

\section{Studies of $CP$-Violation in $b\to c \bar{c} d$ Processes}

Neutral $B$ decays to $CP$-eigenstates that proceed by $b\to c\bar{c}
d$ processes are expected to have the same $CP$-violation as
$b\to c \bar{c} s$ since both are sensitive to the phase of
$B-\bar{B}$ mixing. A small deviation from this expectation is possible
 because of the contribution of $b\to d$ penguin diagrams
(a.k.a. ``penguin pollution'') in the decay modes that are examined.
Penguin pollution may also give
rise to direct $CP$-violation and a $CP$-violating term with a 
$\cos(\Delta m_d \Delta t)$ dependence.

The $b\to c \bar{c} d$ decay modes that have been used so far 
for $CP$-violation studies are $B\to D^{*+} D^{*-}$, $B\to D^{*+} D^-$,
and $B\to J/\psi
\pi^0$\rlap.\,\cite{bib:BaBar_psipi0}$^-$\cite{bib:BaBar_DstDplus} 
The effect of penguin pollution might be expected
to be the largest in $B\to J/\psi \pi^0$ because the penguin contribution
is not color-suppressed in that mode.

For $B\to \psi\pi^0$, with 81 fb$^{-1}$ BaBar 
has a signal of $40\pm 7$ events\cite{bib:BaBar_psipi0} and finds 
\begin{equation}
 \sin 2 \phi_{1eff}(B\to \psi\pi^0) = 0.05 \pm 0.45 \pm 0.16.
\end{equation} 
The corresponding result from Belle is based
on 140 fb$^{-1}$ and uses $89\pm 10$ events\rlap.\,\cite{bib:Belle_psipi0} 
They obtain
\begin{equation}
 \sin 2 \phi_{1eff}(B\to \psi\pi^0) = 0.72^{+0.37}_{-0.42} \pm 0.08.
\end{equation} 
In both cases, the systematic error includes the possibility
of $CP$-violation in a small component of the background that peaks
under the signal.

The $b\to c \bar{c} d$ 
mode $B\to D^{*+} D^{*-}$ has a vector-vector final state
and requires special treatment since it includes contributions
from both $CP$-even and odd components. To extract the $CP$-odd fraction,
one fits the angular distribution in the transversity basis.
The result from BaBar based on a sample with $156\pm 14$ signal
events is,
\begin{equation}
 R_{\perp} = 0.063 \pm 0.055 \pm 0.009,
\end{equation} 
where the quantity $R_{\perp}$ is the fraction of the $CP$-odd component.
The measurement indicates that $B^0\to D^{*+} D^{*-}$ is mostly
$CP$-even.

\begin{figure}
\center
\psfig{figure=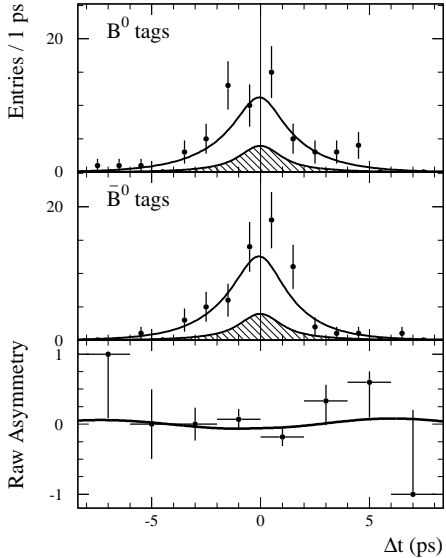,width=6.0truecm}
\caption{BaBar results on $CP$-violation in $B\to D^{*+}D^{*-}$.
The top two figures show the $\Delta t$ distributions
for $B^0$ and $\bar{B}^0$ tags. The third plot shows
the raw time asymmetry distribution.}
\vspace{-0.4cm}
\label{fig:babar_dstdst}
\end{figure}

The time distributions from BaBar for $B\to D^{*+} D^{*-}$
are shown in Fig.~\ref{fig:babar_dstdst}.
 BaBar finds
\begin{equation}
 \sin 2\phi_{1eff} (B\to D^{*+} D^{*-}) = -0.05 \pm 0.29 \pm 0.10,
\end{equation} 
which is about $2.5\sigma$ from the result in $b\to c\bar{c} s$
modes. This may
be a statistical fluctuation or could be an indication that the
Standard Model 
$b\to d$ penguin contribution is large. 
The fit includes the possibility of direct
$CP$-violation. The parameter $\lambda$ is found to be $0.75\pm 0.19\pm 0.02$, 
which is consistent with unity, as expected for no direct $CP$-violation.

Since $B^0\to D^{*+} D^-$ and its charge conjugate are not $CP$-eigenstates, 
a modified treatment is required. There are four rather
than two $CP$-violating observables that are determined from a
time-dependent fit to the different $D^* D$ charge states.

BaBar finds,
\begin{eqnarray}
 S_{+-} = -0.82 \pm 0.75 \pm 0.14, \\
 S_{-+} = -0.24 \pm 0.69 \pm 0.12, \\
 A_{+-} = +0.47 \pm 0.40 \pm 0.12, \\
 A_{-+} = +0.22 \pm 0.37 \pm 0.10.
\end{eqnarray} 
In the limit of no penguins and assuming factorization
in these hadronic decays, $S_{-+}=S_{+-}=-\sin 2\phi_1$ and
$A_{+-}=A_{-+}=0$. The above results for $CPV$ in $B\to D^* D$ decays
are consistent with this limit.

Observation of the $CP$-eigenstate mode $B\to D^+ D^-$ was reported by Belle
at this conference. With 140 fb$^{-1}$, the $5\sigma$ signal contains
$24.3\pm 6.0$ events. In the future, this mode can also be used
for time-dependent measurements of $CPV$ in $b\to c \bar{c} d$
processes.

\begin{figure}
\center
\psfig{figure=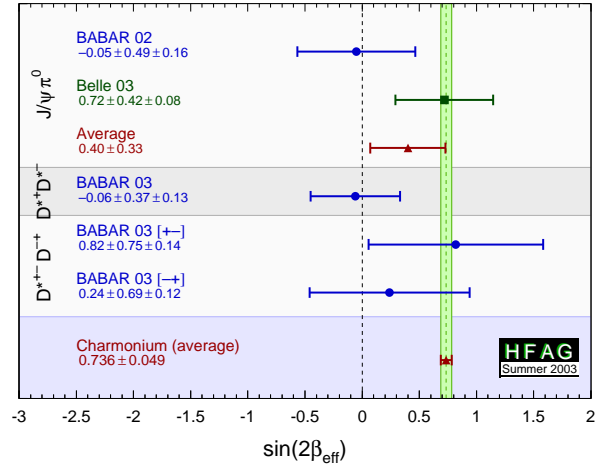,width=8.0truecm}
\caption{Summary plot of results on $CP$-violation 
in $b\to c \bar{c}d$ modes.}
\label{fig:sin2b_ccd}
\end{figure}

The results of $CP$-violation measurements for $b\to c \bar{c} d$
decays are  summarized in Fig.~\ref{fig:sin2b_ccd}. The measurements
are not yet precise enough to definitively demonstrate the presence of
penguin pollution.

\section{Status of $CP$-Violation in $b\to s q \bar{q}$ Penguin
Processes}

In addition to the program of measuring the other remaining 
angles of the unitarity triangle that is discussed
in the contribution by Jawahery\rlap,\,\cite{jawahery}
there is also the question of 
whether there are additional $CP$-violating phases from new
interactions or physics beyond the Standard Model. 
At the moment, such new phases are
poorly constrained. 

One way to attack this question is
to measure the time-dependent $CP$-asymmetry in penguin-dominated
modes such as $B^0\to \phi K_S^0$, $B^0\to \eta^{'} K_S^0$
or $B^0\to K_S^0\pi^0$, where heavy
new particles may contribute inside the loop, 
and compare it to the asymmetry in $B^0\to J/\psi K_S^0$ and related
$b\to c\bar{c} s$ charmonium modes.

\begin{figure}[htb]
\center
%\rule{5cm}{0.2mm}\hfill\rule{5cm}{0.2mm}
%\vskip 4cm
%\rule{5cm}{0.2mm}\hfill\rule{5cm}{0.2mm}
\psfig{figure=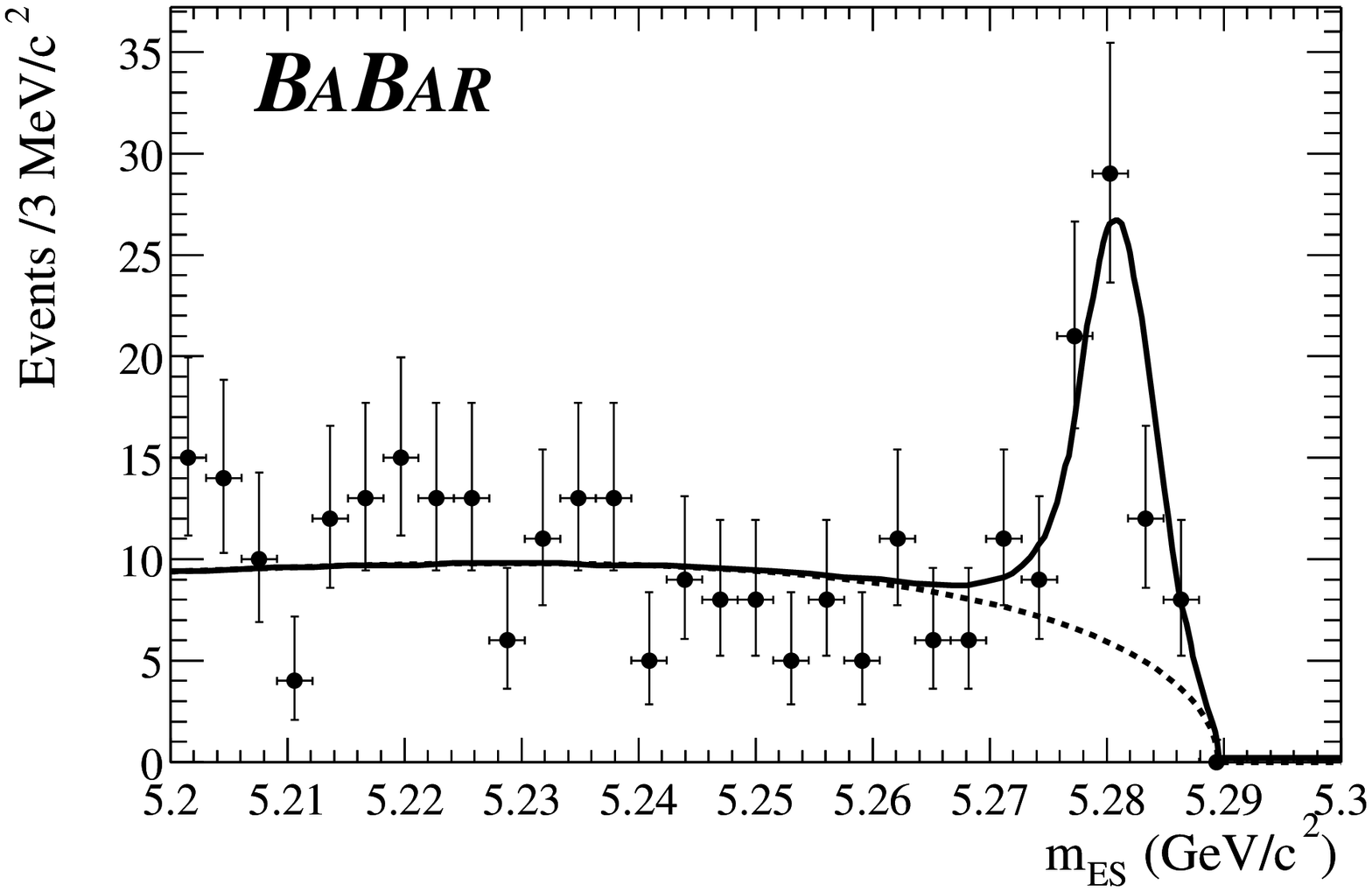,width=1.5truein,height=1.5truein}
\psfig{figure=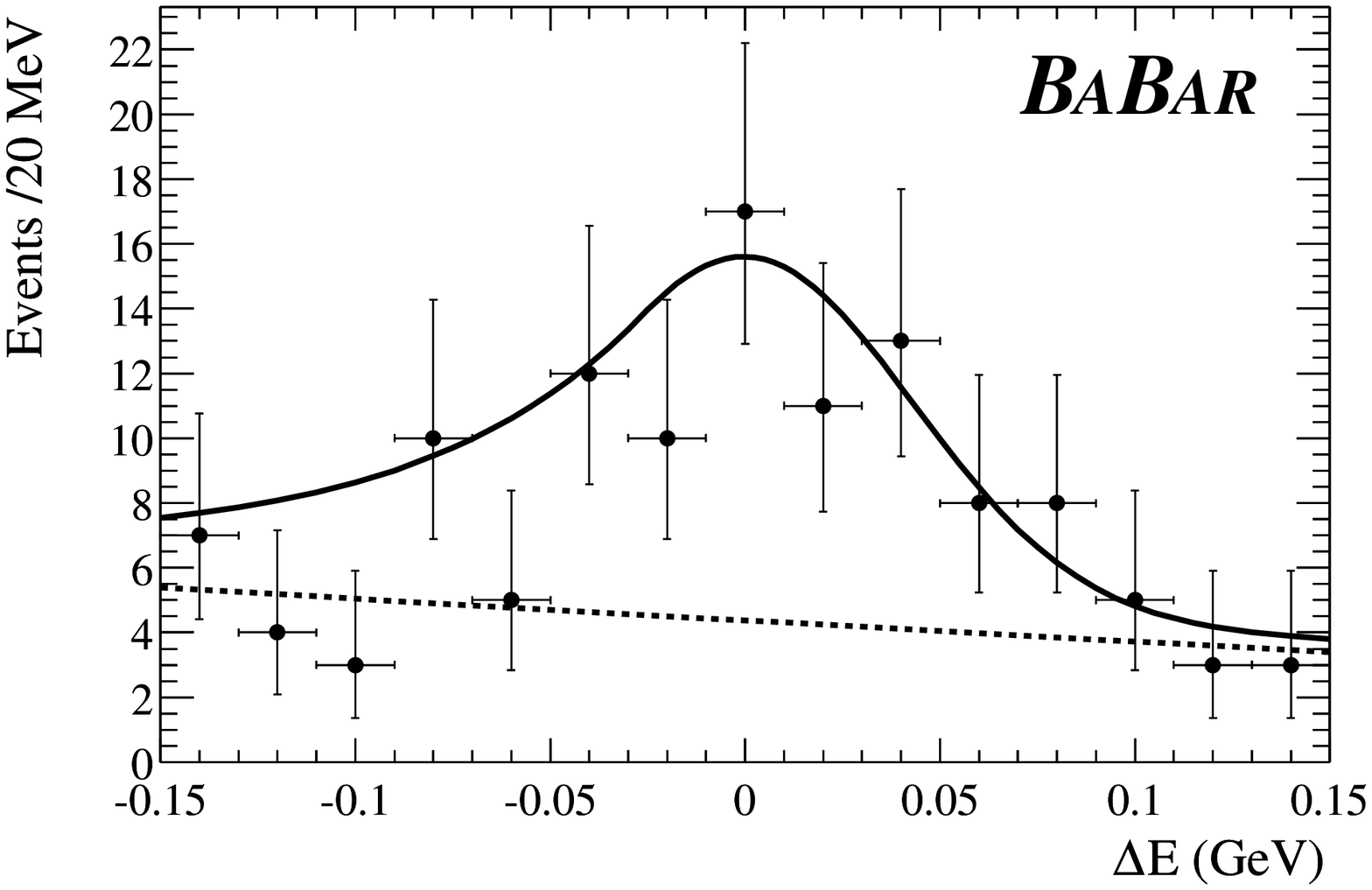,width=1.5truein,height=1.5truein}
\caption{Beam constrained mass 
and $\Delta E$ distributions for $B\to K_S^0\pi^0$ from BaBar.}
\label{fig:mbc_kspi0} 
\end{figure}

\begin{figure}[htb]
\center
\psfig{figure=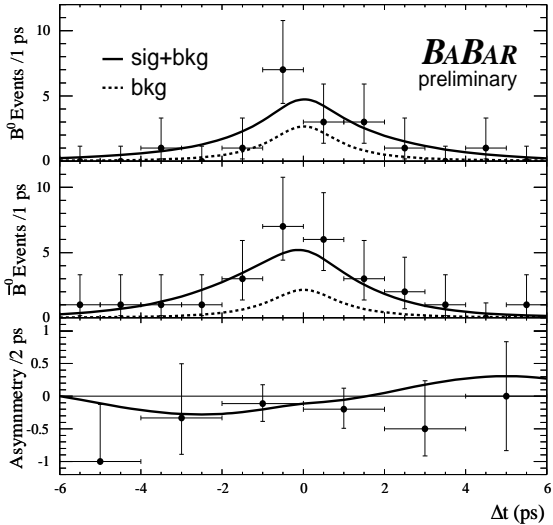,width=3.0truein,height=3.0truein}
\caption{BaBar data on $B\to K_S^0\pi^0$.
The top two figures show the $\Delta t$ distributions
for $B^0$ and $\bar{B}^0$ tags, separately. The third plot shows
the raw time asymmetry distribution.}
\label{fig:babar_kspi0_asym} 
\end{figure}

The mode $B\to K_S\pi^0$ proceeds through a $b\to s d\bar{d}$ transition. 
The BaBar data on $B\to K_S^0\pi^0$ are shown in Fig.~\ref{fig:mbc_kspi0}.
To be useful for time-dependent $CPV$ studies
 at least one of charged pions from the
$K_S^0$ must be detected in the 
BaBar silicon vertex detector\rlap.\,\cite{bib:BaBar_sss} 
There are $123\pm 16$ events of this type that are then used to
obtain 
\begin{equation}
 \sin 2 \phi_{1eff}(B\to K_S^0\pi^0) = 0.48^{+0.38}_{-0.47}\pm 0.11.
\end{equation} 
The time distributions are shown in Fig.~\ref{fig:babar_kspi0_asym}.
The direct $CP$-violation parameter is 
$A=-0.40^{+0.28}_{-0.27}\pm 0.10$\rlap.\,\cite{bib:BaBar_sss}
 When A is fixed to zero, the value of $S=\sin(2\phi_{1eff})$ 
shifts slightly to $0.41^{+0.41}_{-0.48}\pm 0.11$.
The results for $B\to K_S\pi^0$ are consistent with the value
from the $b\to c \bar{c} s$ modes, $\sin 2\phi_1= 0.736\pm 0.049$.  

\begin{figure}[htb]
\center
%\rule{5cm}{0.2mm}\hfill\rule{5cm}{0.2mm}
%\vskip 4cm
%\rule{5cm}{0.2mm}\hfill\rule{5cm}{0.2mm}
\psfig{figure=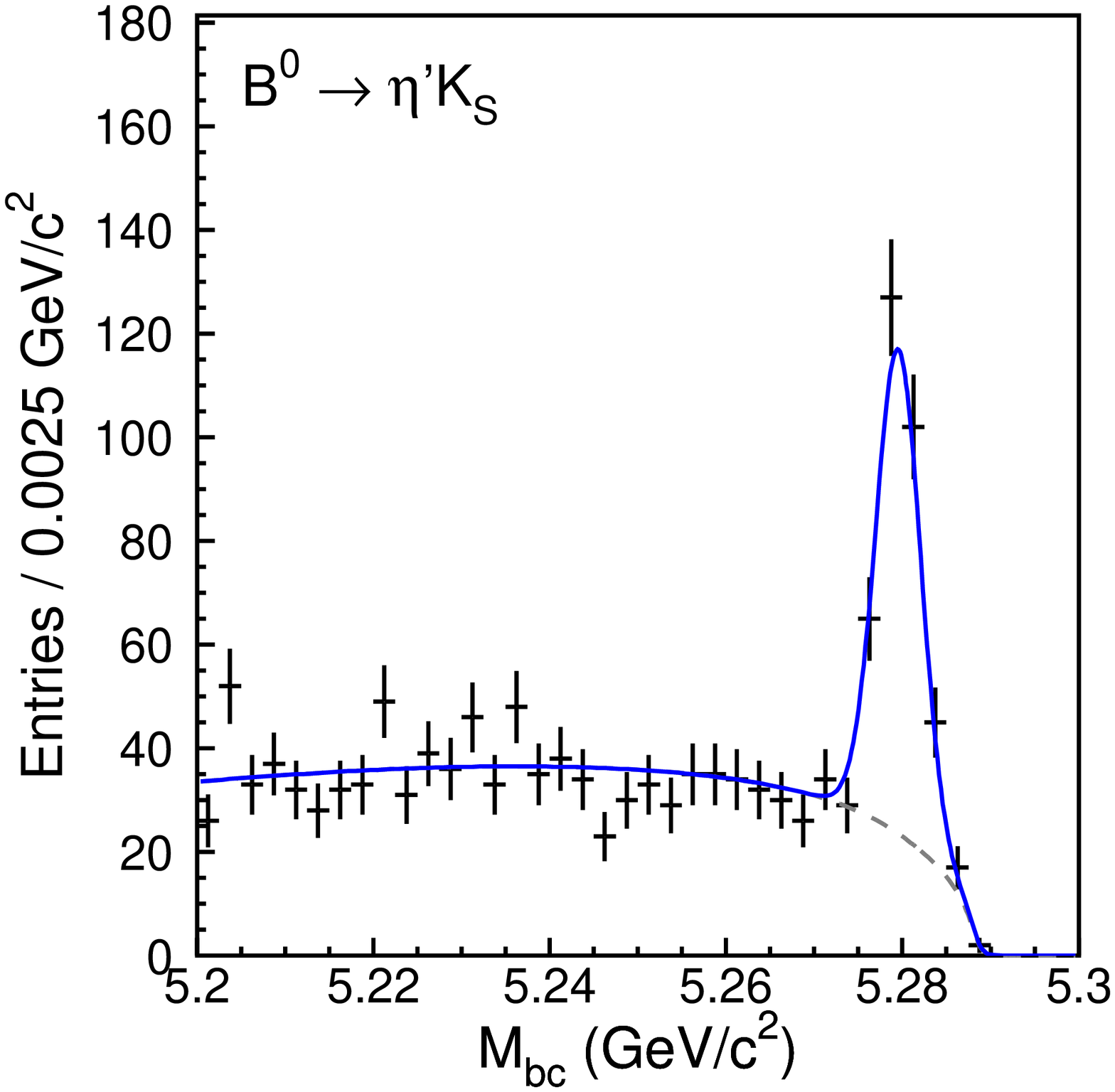,width=1.5truein,height=1.5truein}
\psfig{figure=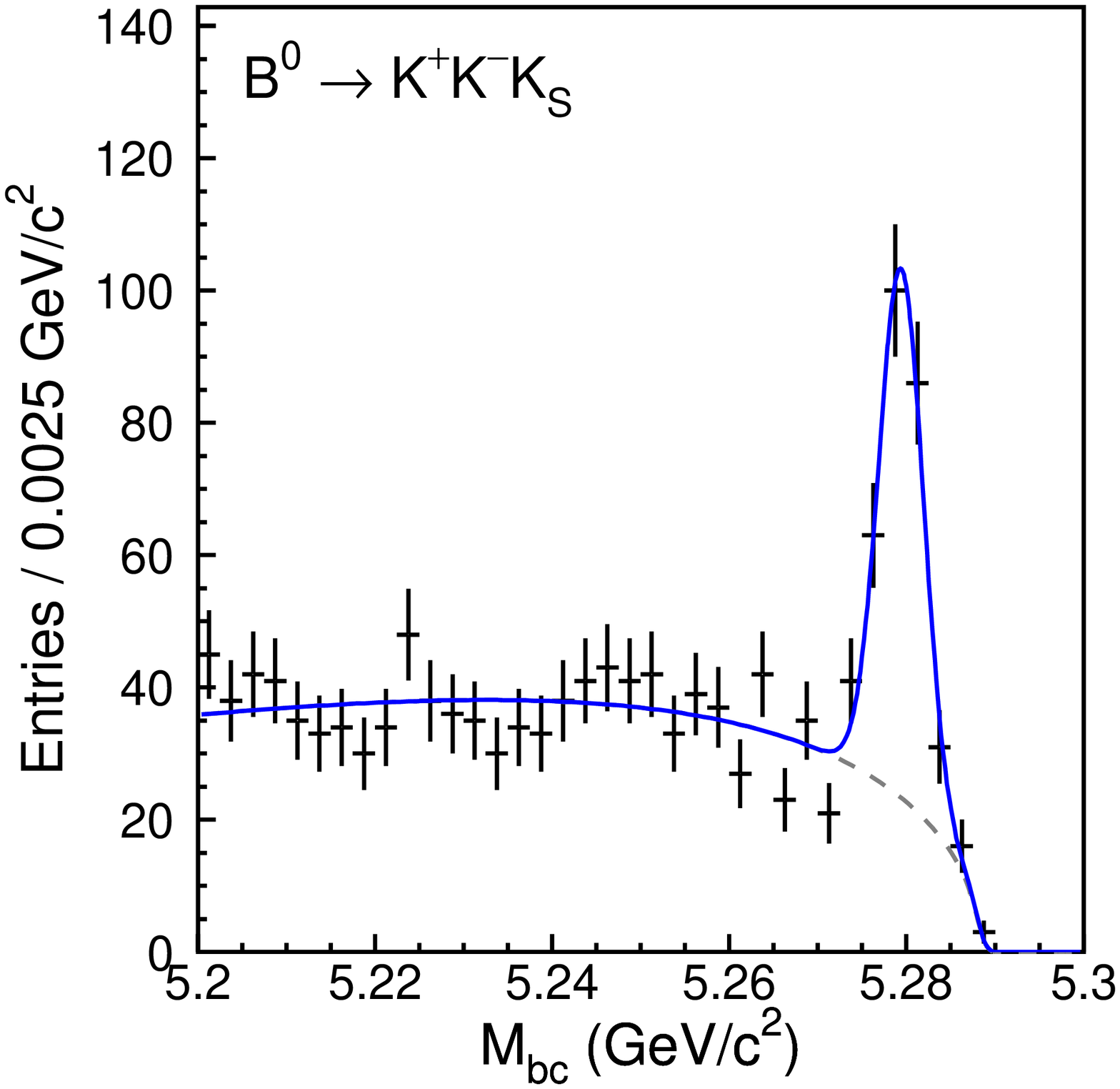,width=1.5truein,height=1.5truein}
\caption{Beam constrained mass distributions for $B\to \eta^{\prime} K_S^0$
 (left) and $B\to K^+ K^- K_S^0$ (right).}
\label{fig:mbc_other} 
\end{figure}

\begin{figure}[htb]
\center
\psfig{figure=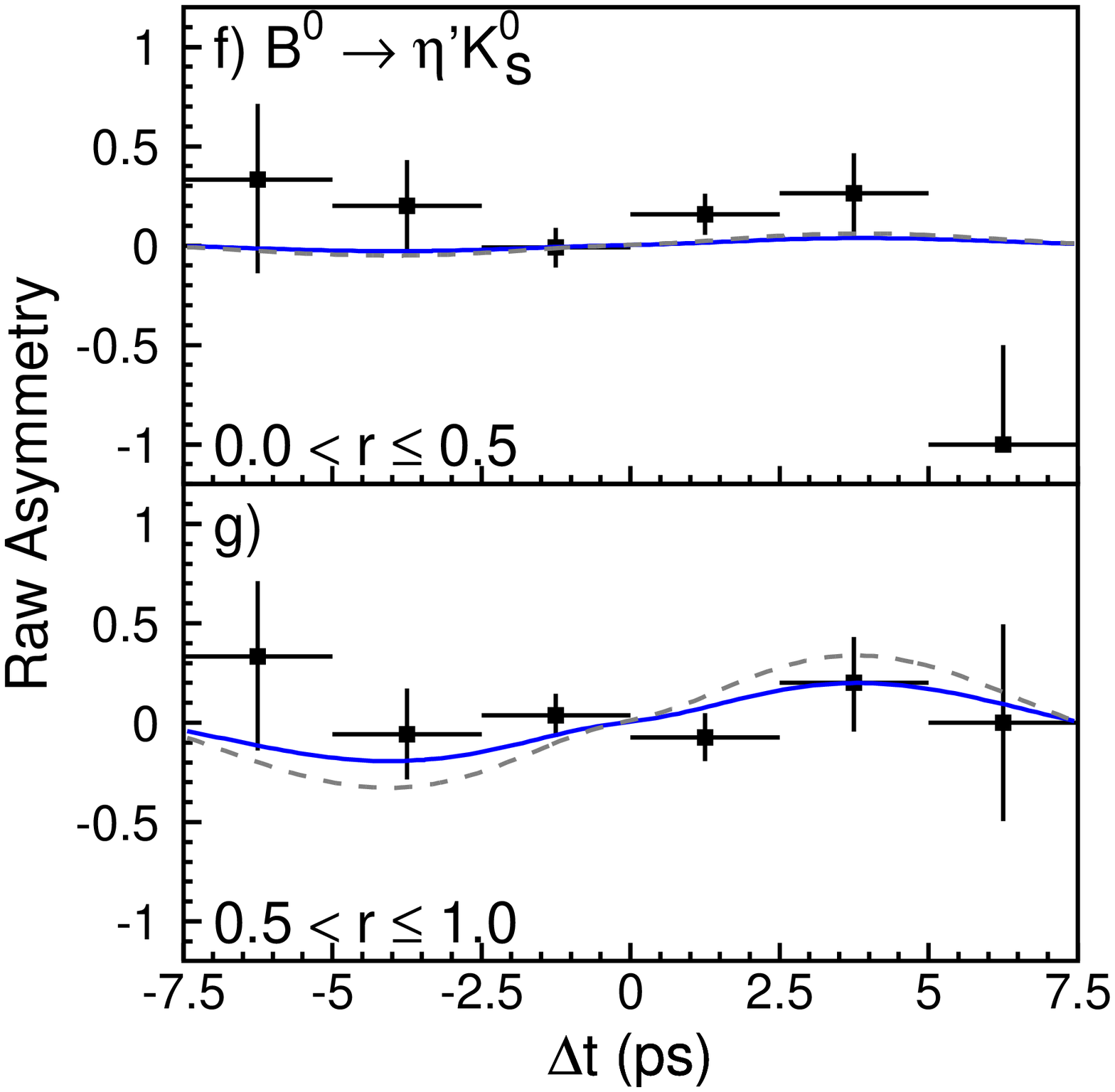,width=2.5truein,height=2.5truein}
\caption{Belle data for the raw asymmetry in $B^0\to \eta^{\prime} K_S^0$.
The upper plot shows the data for low-quality tags
while the lower plot shows the higher quality tags.
The dashed curves are the expectations from the Standard Model.}
\label{fig:belle_etapks_asym} 
\end{figure}

\begin{figure}[htb]
\center
\psfig{figure=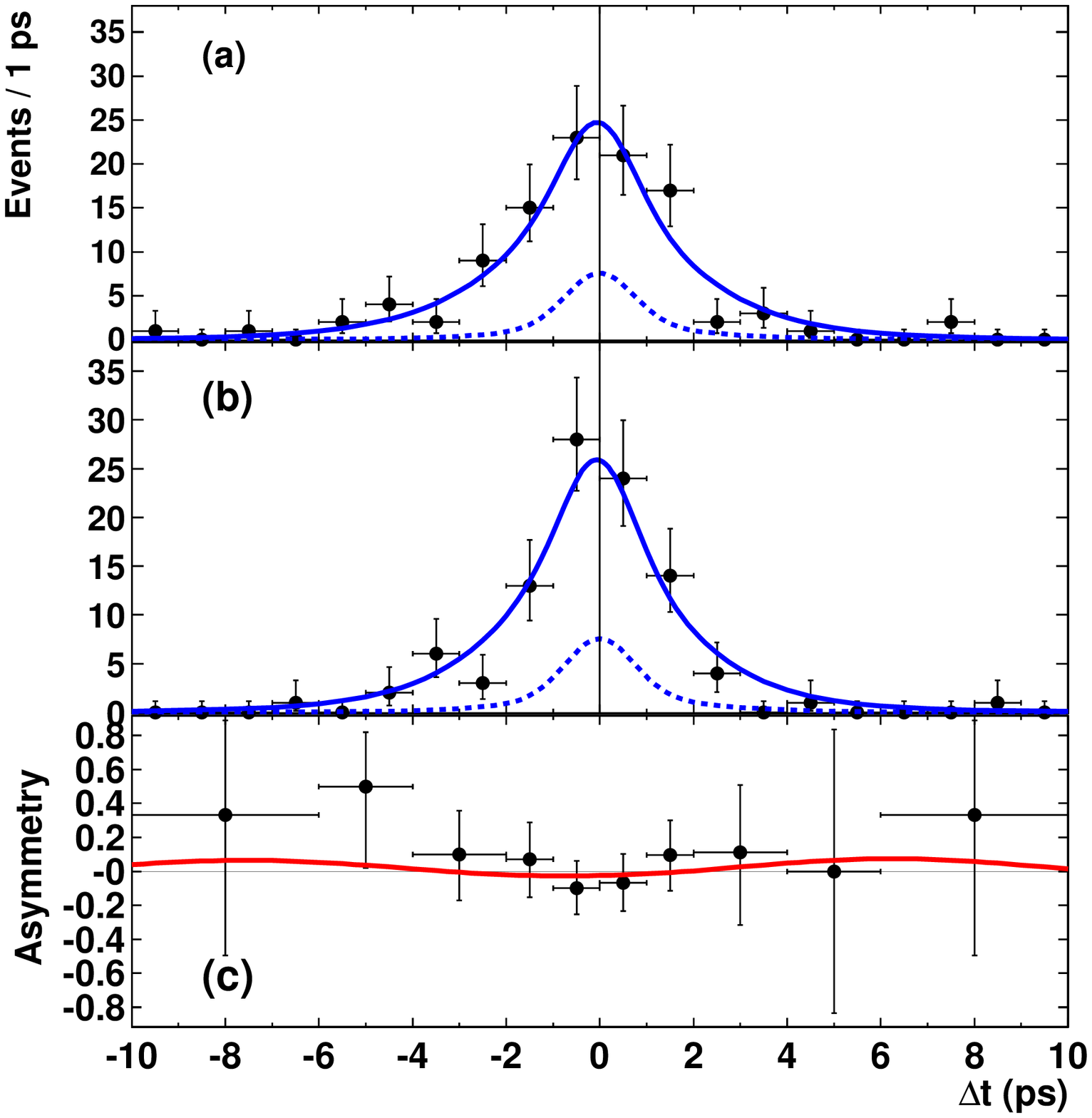,width=3.0truein,height=3.0truein}
\caption{BaBar data on $B\to \eta^{\prime} K_S^0$.
The top two figures show the $\Delta t$ distributions
for $B^0$ and $\bar{B}^0$ tags, separately. The third plot shows
the raw time asymmetry distribution.}
\label{fig:babar_etapks_asym} 
\end{figure}

The mode $B\to \eta^{\prime} K_S^0$ is expected to include
contributions from $b\to s \bar{u} u$ and $b\to s\bar{d} d$ 
penguin processes. 
The beam constrained mass distribution for the $B\to \eta^{\prime} 
K_S^0$ sample used by Belle is shown in Fig.~\ref{fig:mbc_other}
and contains $244\pm 21$ signal events\rlap.\,\cite{bib:Belle_etapks}
Belle finds (Fig.~\ref{fig:belle_etapks_asym}),
\begin{equation}
 \sin 2 \phi_{1eff}(B\to \eta^{\prime} K_S^0) = 0.43\pm 0.27\pm 0.05
\end{equation} 
The BaBar data is shown in Fig.~\ref{fig:babar_etapks_asym}. They obtain,
\begin{equation}
 \sin 2 \phi_{1eff}(B\to \eta^{\prime} K_S^0) = 0.02\pm 0.34\pm 0.03
\end{equation} 
The average of these two results for $B\to \eta^{\prime} K_S^0$
is about 2.2$\sigma$ from the $b\to
c\bar{c} s$ measurement, which is the Standard Model expectation.

The decay mode $B\to K^+ K^- K_S^0$, where $K^+ K^-$ combinations
consistent with the $\phi$ have been removed, is found by Belle to
be dominately $CP$-odd\cite{bib:Garmash} 
and thus can be treated as a $CP$-eigenstate
and used for studies of time-dependent
$CP$-violation in $b\to s q \bar{q}$ processes. 
The beam constrained mass distribution for the $B\to K^+ K^- K_S^0$ sample
used by Belle is shown in Fig.~\ref{fig:mbc_other}. There are $199\pm
18$ signal events.
Belle obtains,
\begin{equation}
 \sin 2 \phi_{1eff}(B\to K^+ K^-K_S^0) = 0.51\pm 0.26\pm
0.05^{+0.18}_{-0.00},
\end{equation} 
where the third error is due to the uncertainty in the $CP$ content
of this final state\rlap.\,\cite{bib:Garmash}
The results for $B\to K^+ K^- K_S^0$ are also consistent with
$b\to c\bar{c} s$ decays. However, in this decay there is also
the possibility of ``tree-pollution'', the contribution of 
the $b\to u \bar{u} s$ tree amplitude that may complicate the
interpretation of the results\rlap.\,\cite{bib:ligeti}

The $B^0\to \phi K_S^0$ decay, 
which is dominated by the $b \to s\bar{s} s$
transition, is an especially unambiguous and
sensitive probe of new $CP$-violating phases from 
physics beyond the SM\rlap.\,\cite{bib:lucy}
The SM predicts that measurements of $CP$-violation in this
mode should yield $\sin 2\phi_1$ to
a very good approximation\rlap.\,\cite{bib:tree-penguin,bib:ligeti}
A significant deviation in the time-dependent $CP$-asymmetry in
this mode from what is observed
in $b \to c\overline{c}s$ decays would be evidence for a new
$CP$-violating phase.

\begin{figure}[htb]
\center
\psfig{figure=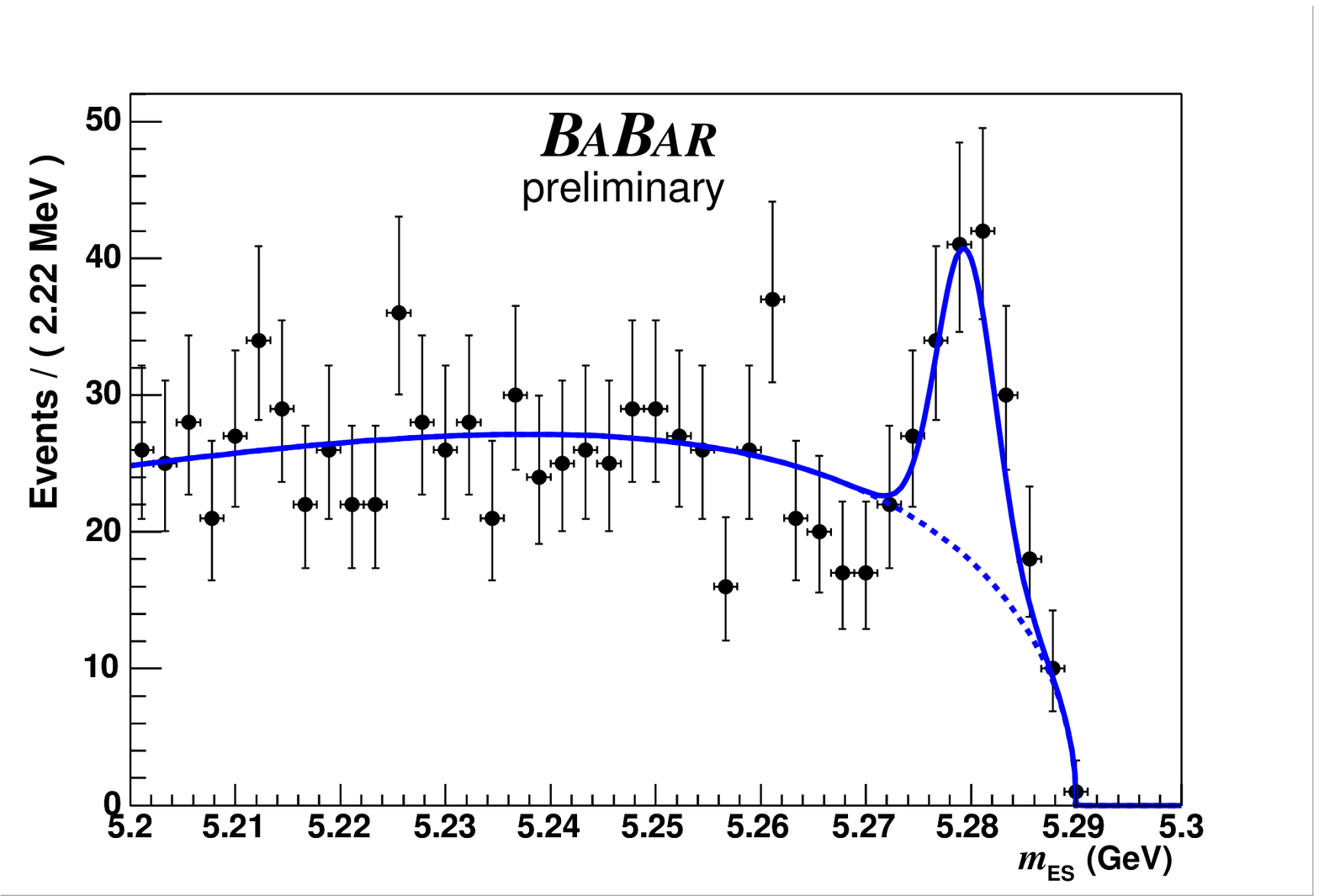,width=1.5truein,height=1.5truein}
\psfig{figure=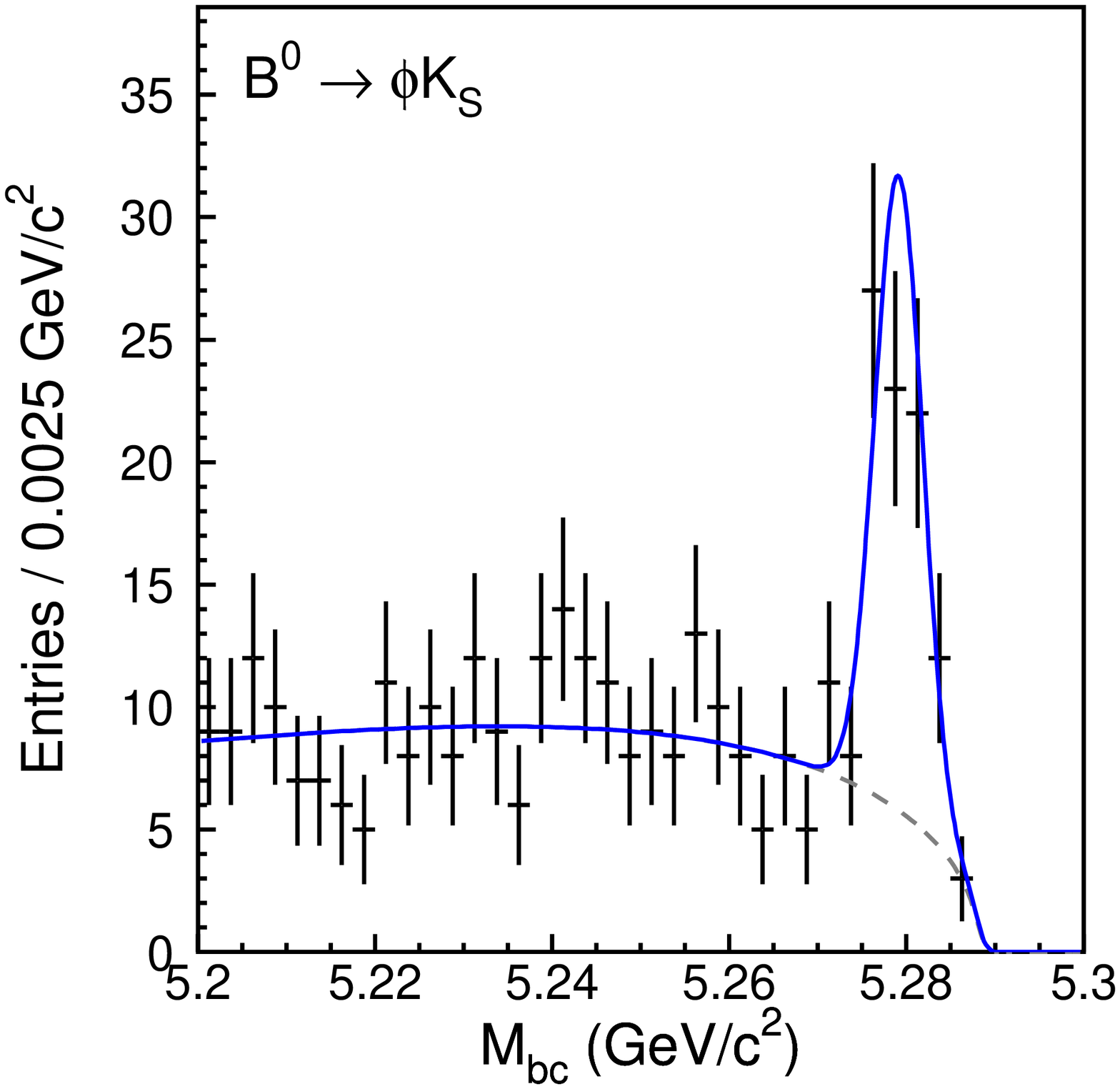,width=1.5truein,height=1.5truein}
\caption{Beam constrained mass distributions for $B\to \phi K_S^0$
from BaBar (left) and Belle(right).}
\label{fig:mbc_phiks} 
\end{figure}

\begin{figure}[htb]
\center
\psfig{figure=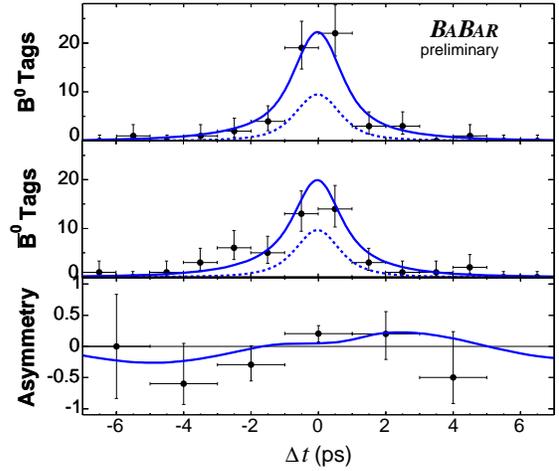,width=3.0truein}
\caption{BaBar time difference and asymmetry data distributions in
$B\to \phi K_S^0$.
The top two figures show the $\Delta t$ distributions
for $B^0$ and $\bar{B}^0$ tags, separately. The third plot shows
the raw time asymmetry distribution.}
\label{fig:babar_dtphiks}
\end{figure}

The $B\to\phi K_S^0$ sample used by BaBar is shown in
Fig.~\ref{fig:mbc_phiks}. The signal, obtained from
a sample with an integrated luminosity of 110 fb$^{-1}$,
contains $70\pm 9$ events\rlap.\,\cite{bib:BaBar_sss}
The time distributions for the
BaBar data are shown in Fig.~\ref{fig:babar_dtphiks}.
They obtain
\begin{equation}
 \sin 2 \phi_{1eff}(B\to \phi K_S^0) = 0.45 \pm 0.43 \pm 0.07.
\end{equation} 
This value is consistent with the Standard Model expectation, but is
somewhat different from the value obtained
with the 81 fb$^{-1}$ sample, which was
$\sin 2 \phi_{1eff}=-0.18\pm 0.51 \pm 0.09$. 
The new result includes more data
and a reprocessing of the old data sample. After extensive checks
with data and Toy Monte Carlo studies, the large change 
in the central value is attributed
to a $1\sigma$ statistical fluctuation\rlap.\,\cite{lancieri}

\begin{figure}[htb]
\center
\psfig{figure=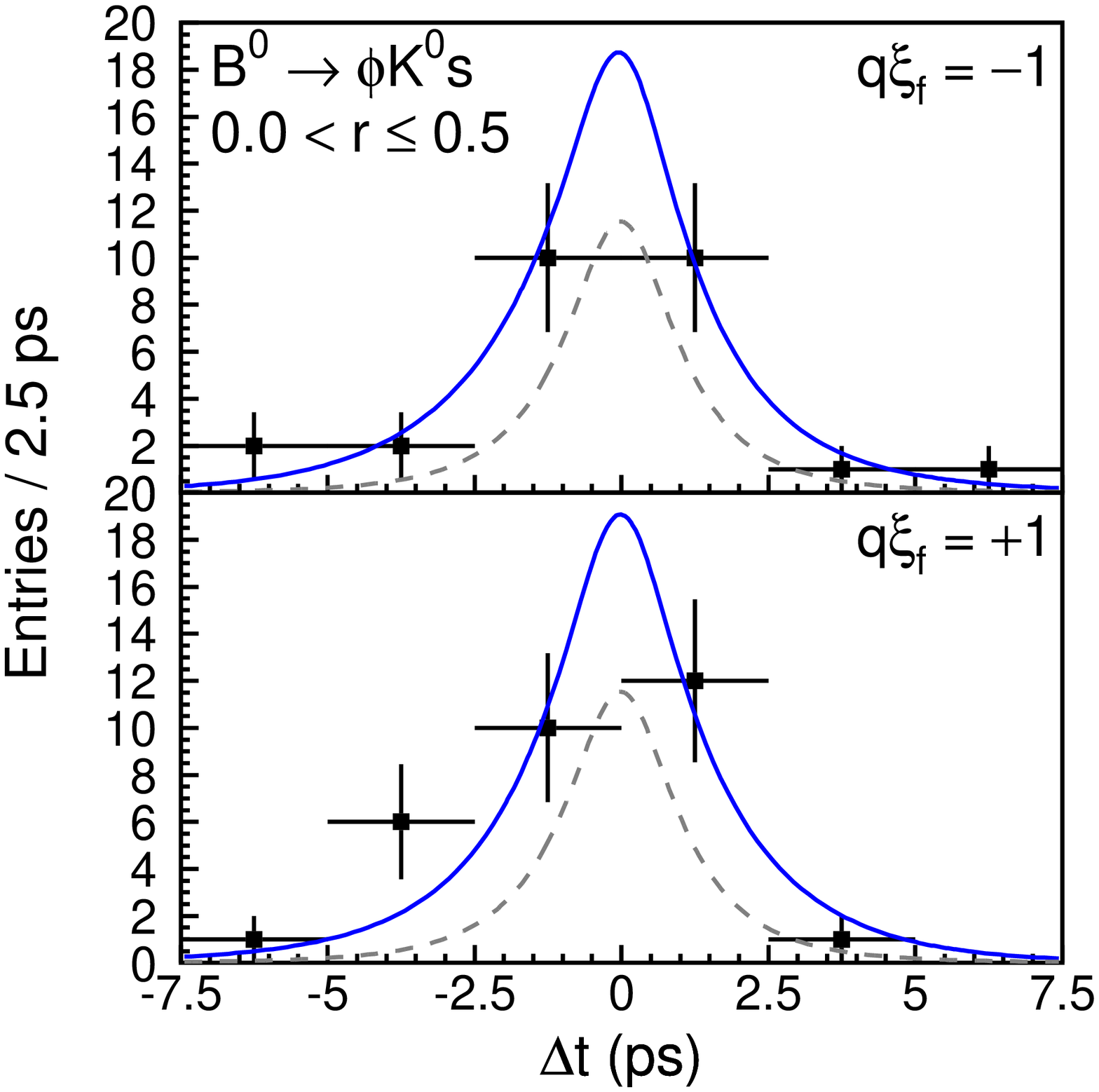,width=1.5truein}
\psfig{figure=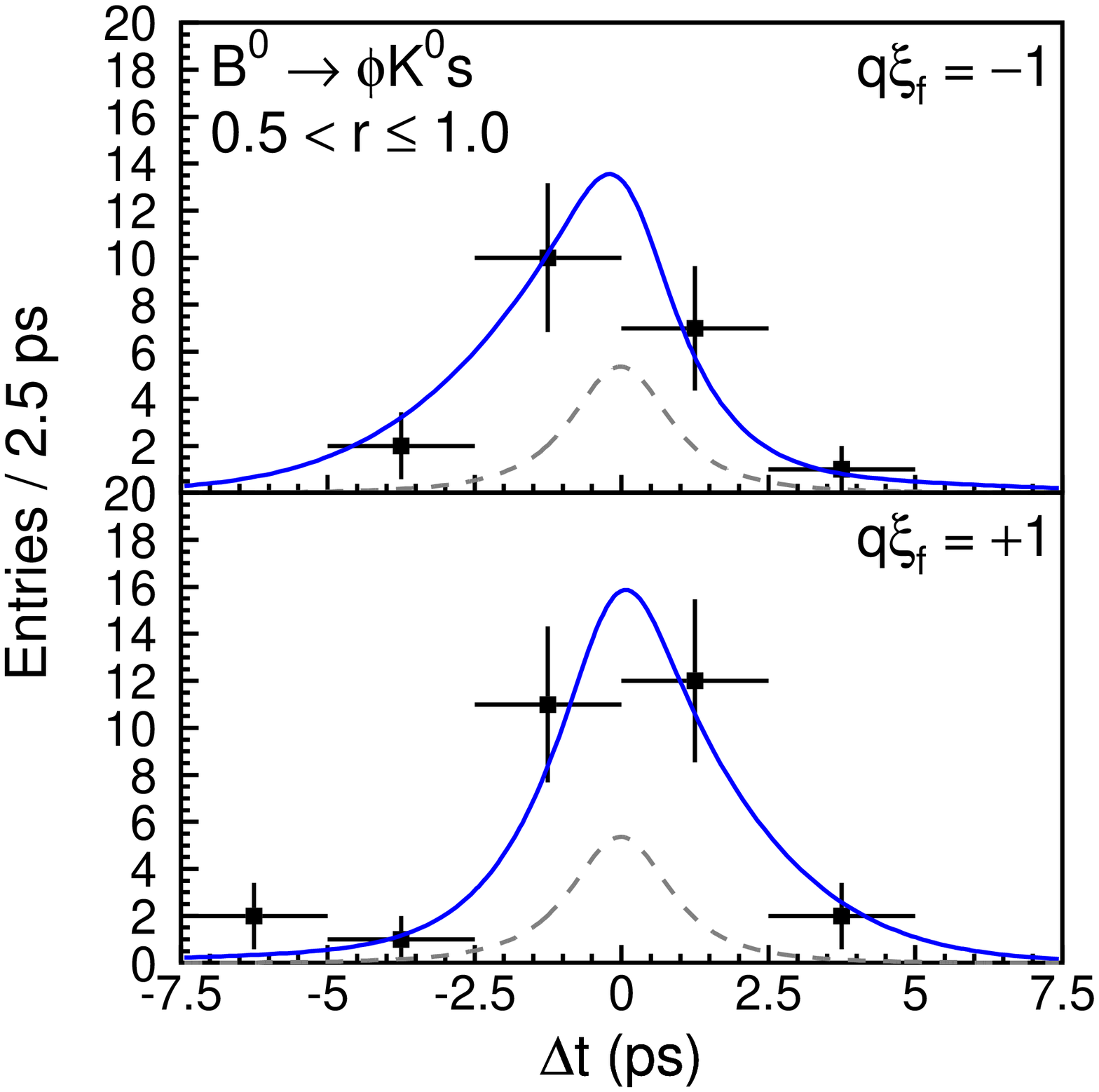,width=1.5truein}
\caption{Belle data: (left) $\Delta t$ distributions 
for low-quality tags and (right) for high-quality tags.
The dashed curves show the background contributions.}
\label{fig:dtphiks_belle}
\end{figure}

\begin{figure}[htb]
\center
\psfig{figure=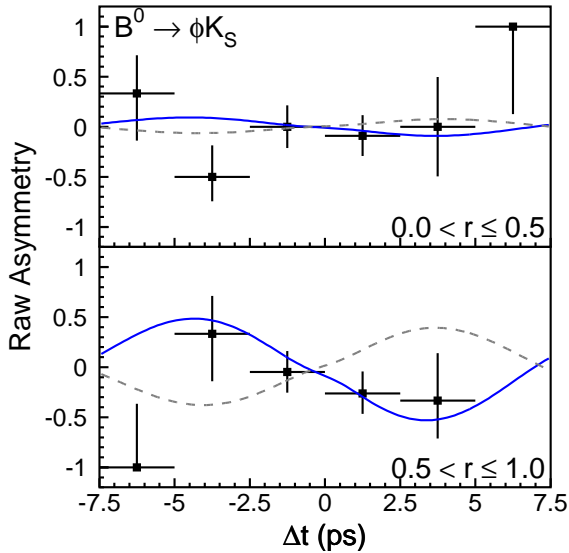,width=8.0truecm}
\caption{Belle data for the raw asymmetry in $B^0\to \phi K_s^0$.
The upper plot shows the data for low-quality tags
while the lower plot shows the higher quality tags.
The dashed line is the expectation from the Standard Model.}
\label{fig:asym_phiks_belle}
\end{figure}

The $B\to\phi K_S^0$ sample used by Belle is shown in the right panel 
of Fig.~\ref{fig:mbc_phiks}. The selection criteria are described in
 detail elsewhere\rlap.\,\cite{bib:Belle_phik,bib:Belle_phiks_2003}
The signal contains $68\pm 11$ events.
Figure~\ref{fig:asym_phiks_belle} shows the raw asymmetries 
from Belle in two regions of the flavor-tagging
parameter $r$. While the numbers of events in the two regions are similar,
the effective tagging efficiency is much larger 
and the background dilution is smaller in the region $0.5 < r \le 1.0$.
The solid curves show the results of the unbinned maximum-likelihood 
fit to the $\Delta t$ distribution.

The observed $CP$-asymmetry for $B^0 \to \phi K_S^0$
in the region $0.5 < r \le 1.0$ (Fig.~\ref{fig:asym_phiks_belle} (lower panel))
indicates the difference from the SM expectation (dashed curve). 
Note that these projections onto the $\Delta t$ axis do not take into
account event-by-event information (such as the signal fraction, the
wrong tag fraction and the vertex resolution) that is used in the
unbinned maximum likelihood fit.

The contamination of $K^+ K^- K_S^0$ events in the $\phi K_S^0$ sample 
($7.2\pm1.7$\%) is small.
Finally, backgrounds from the $B^0 \rightarrow f_0(980) K_S^0$ decay, 
which has the opposite 
$CP$-eigenvalue to $\phi K_S^0$, are found to be small
($1.6^{+1.9}_{-1.5}$\%).  The influence of these backgrounds
is treated as a source of systematic uncertainty.

Belle obtains
\begin{equation}
 \sin 2 \phi_{1eff}(B\to \phi K_S^0) = -0.96 \pm 0.5^{+0.09}_{-0.11}
\end{equation} 
from their likelihood fit to the $\phi K_S^0$ data.
The likelihood function is parabolic and well-behaved.
An evaluation of the significance of the result using the
Feldman-Cousins method and allowing for systematic uncertainties
shows that this result deviates by 3.5$\sigma$ from the Standard
Model expectation\rlap.\,\cite{bib:Belle_phiks_2003}

\begin{figure}[htb]
\center
\psfig{figure=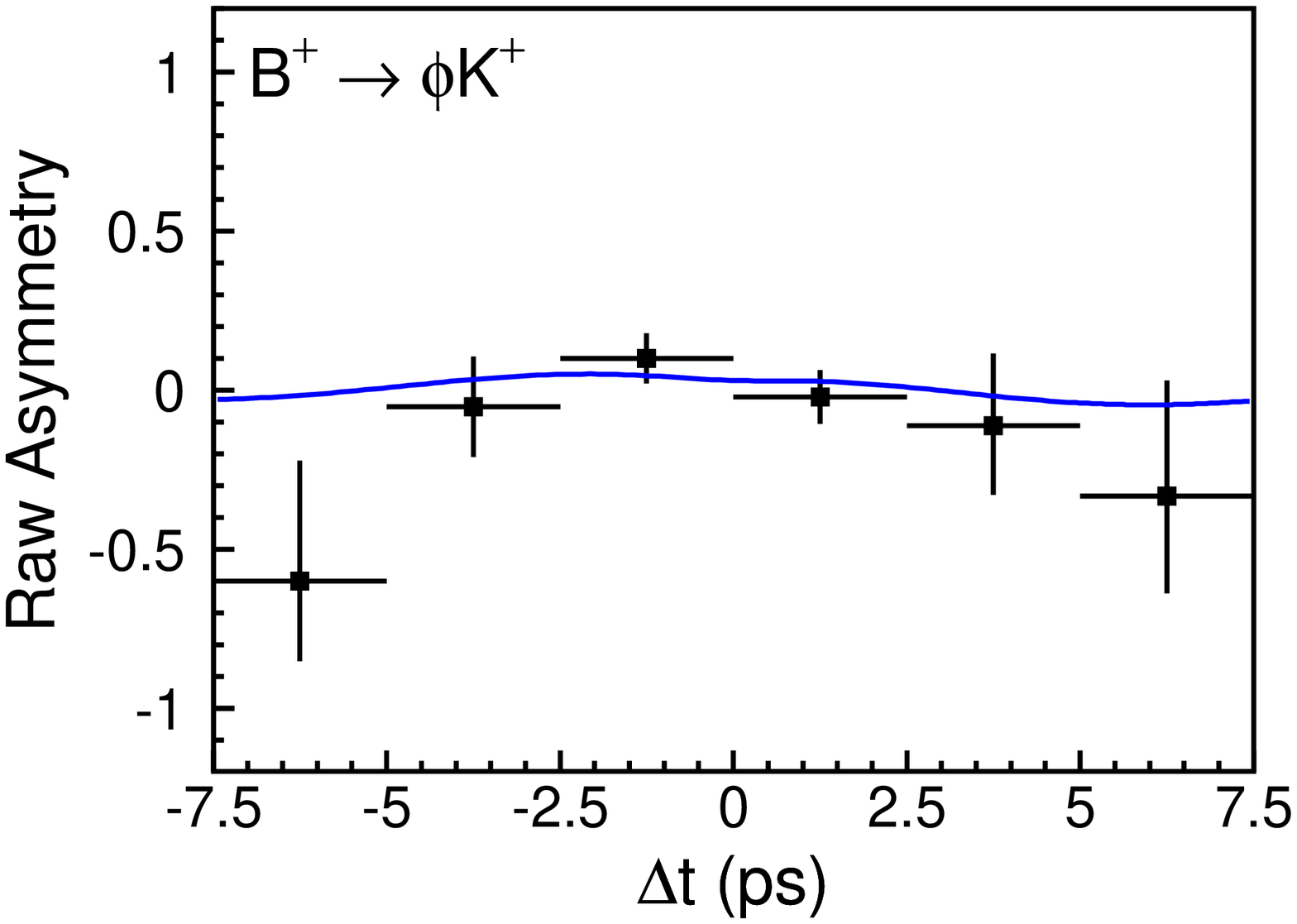,width=1.5truein}
\psfig{figure=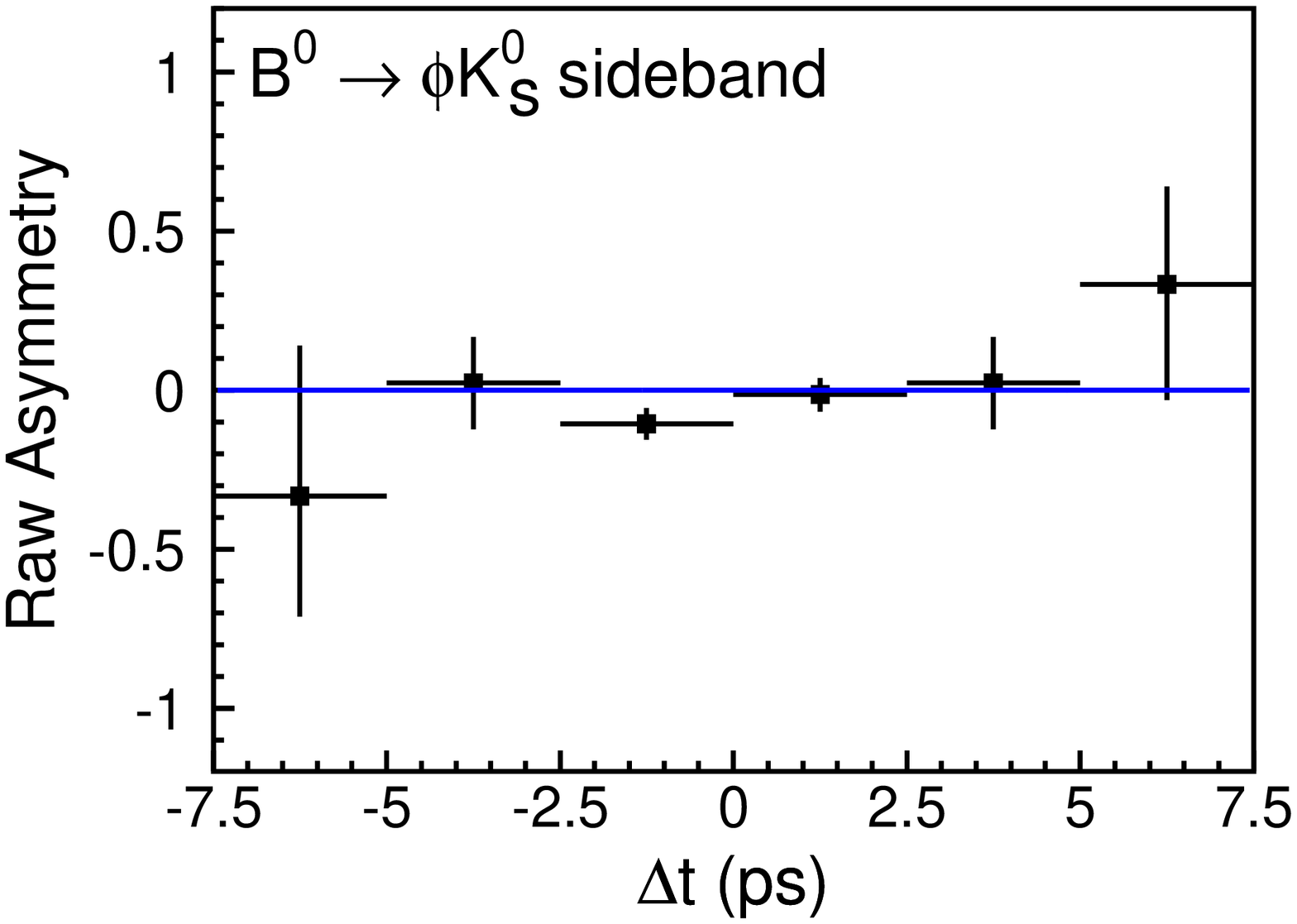,width=1.5truein}
\caption{Belle data: consistency checks of the $B\to \phi K_S^0$
analysis. The asymmetries in (a)  the $B^{\pm}\to \phi K^{\pm}$ sample
 and (b) the $B\to\phi K_S^0$ sideband sample.}
\label{fig:dtphiks_checks}
\end{figure}

The Belle group performed a number of validation checks
for their $B\to \phi K_S^0$ $CP$-violation result.
Fits to the same samples with the 
direct $CP$-violation parameter ${\cal A}$ fixed at zero yield
$sin 2\phi_{1eff} = -0.99\pm0.50$(stat) for $B^0\to \phi K_S^0$.
As a consistency check for the ${\cal S}$ term, the same fit procedure
is applied to the charged 
$B$ meson decays $B^+ \to \phi K^+$.
The result is 
${\cal S} = -0.09\pm0.26$(stat), ${\cal A} = +0.18 \pm 0.20$(stat) 
for $B^+ \to \phi K^+$ decay. 
The results for the ${\cal S}$ term is consistent with no $CP$-asymmetry, 
as expected. The asymmetry distribution is shown in
Fig.~\ref{fig:dtphiks_checks}. 
In addition, the $\phi K_S^0$ sideband
has been examined as shown in Fig.~\ref{fig:dtphiks_checks}. 
No asymmetry is found in that sample.

\begin{figure}
\center
%\rule{2cm}{0.2mm}\hfill \rule{2cm}{0.2mm}
%\vskip 6cm
%\rule{2cm}{0.2mm}\hfill \rule{2cm}{0.2mm} 
\psfig{figure=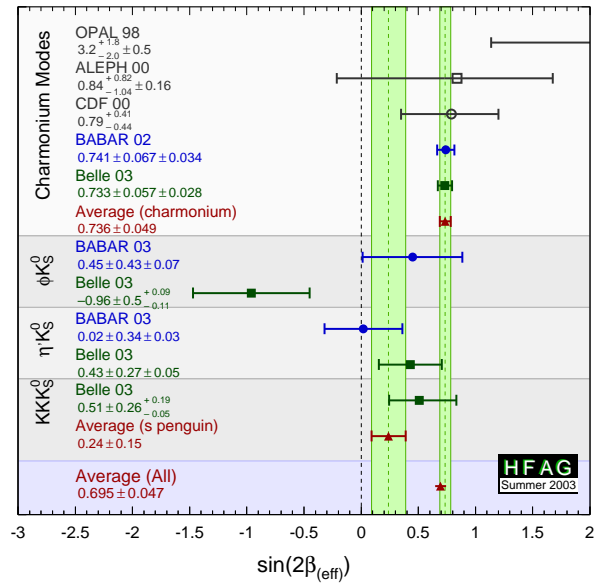,width=8.0truecm}
\caption{Summary plot of results on $\sin 2\phi_1$ and $\sin 2\phi_{1eff}$
in $b\to c\bar{c} s$ and $b\to s \bar{q} q$ modes.}
\label{fig:sin2beta_sum}
\end{figure}

\section{Conclusion}

Belle presented a new measurement of time-dependent $CP$-violation 
in $b\to c\bar{c} s$ $CP$-eigenstates. This result and previous results
from BaBar are in good agreement with each other and with the hypothesis
that the Kobayashi-Maskawa phase is the source of $CP$-violation.

Studies of $CP$-violation in $b\to c\bar{c} d$ modes are progressing.
In $B\to D^{*+} D^{*-}$ decays, BaBar observes a $2.5\sigma$ hint
for penguin pollution. More data and measurements are needed to
clarify whether penguin pollution is present in this class of
decays.

In $B\to\phi K_S^0$ decays there was a surprise. With 140 
fb$^{-1}$ Belle observed
a 3.5$\sigma$ deviation from the Standard Model expectation. 
This could be an indication of new physics from heavy particles
in the $b\to s \bar{s}$ penguin loop. However, BaBar's
value moved closer to the Standard Model with the addition of new
data and reprocessing. More precise measurements of the other 
$b\to s q \bar{q}$ modes can further
constrain phases from new physics. 
For example, new physics may
contribute differently to pseudoscalar-vector 
and pseudoscalar-pseudoscalar modes\rlap.\,\cite{grossman}

The results of $CP$-violation measurements for $b\to s q \bar{q}$
penguin decays are  summarized in Fig.~\ref{fig:sin2beta_sum}. 
The world average for all $b\to s$
penguin decays (shown by the dotted line) appears
to be displaced from the average for $b\to c\bar{c} s$ modes.
The high energy physics community will
require that this experimental issue be resolved conclusively in the future. 
This will require large data samples with integrated luminosities of
at least 1 ab$^{-1}$ or 1000 fb$^{-1}$.

\section*{Acknowledgments}
I thank Harry Cheung for his patience and Andreas Hoecker 
for his contributions
to the figures. I also wish to thank
my colleagues at KEK-B, PEP-II, Belle and BaBar for their
extraordinary contributions to the work reported here.

\balance

%%%%%%%%%%%%%%%%%%%%%%%%%%%%%%%%
%  Question and Answer Section %
%%%%%%%%%%%%%%%%%%%%%%%%%%%%%%%%
% Use clear page to make sure everything is flush and a new
% page is started (not just a new column)
%%%%%%%%%%%%%%%%%%%%%%%%%%%%%%%%
\clearpage
\twocolumn[
\section*{DISCUSSION}
]

\begin{description}

\item[Stefan Spanier] (University of Tennessee): \par
%\vskip 3mm
1) Unfortunately, the plenary session gives
the audience only a limited
chance to help you to establish the results
by asking detailed questions.

2) Knowing the previous value of $S= -0.7\pm 0.6$ from Belle,
the newly added statistics must lead to
an unphysical value of $S<-1.4$ leading typically
to large correlations in S and C
(pathological behavior) in this new sample.
How probable is the value in the new sample?

3) How strong is the $CP$-asymmetry in the background?

\item[Tom Browder{\rm :}] \hspace{4cm}\par
%\vskip 3mm
1) A special breakout session is planned later in the Symposium.

2) For a true value near $S=-1$, the values in the new
sample are quite consistent with Toy Monte Carlo studies.
There is no statistically pathological behaviour in either
old or new data samples. 
The observed errors are actually slightly
larger than expected.

3) The background from $B\to f_0 K_S^0$
and $B\to K^+ K^- K_S^0$ decays is small and the $CP$-asymmetry
from these backgrounds is
included in the systematic error.

\item[Alex Kagan] (Cincinnati): 
Can you show the raw BaBar data for $S(\phi K_S^0)$ again?

\item[Tom Browder{\rm :}]
Yes. Note that a figure with this data was included in the talk and
appears in the Proceedings. 

\item[Hitoshi Murayama] (Berkeley): 
On the $\phi K_S^0$ mode, the change in the BaBar result was
attributed to a statistical fluctuation.
They have added only 40\% more data.
How is that possible?
Do you have a breakdown of the asymmetry between the
previous and new data samples?

\item[Tom Browder{\rm :}]
Not only was more data added, but the old BaBar data sample
was also reprocessed. After reprocessing, a small number of
events changed from $B^0$ tags to
$\bar{B^0}$ tags (or vice versa). This accounts for the 
shift in the central value.

\end{description}


\begin{thebibliography}{99}
\bibitem{bib:CC}
The inclusion of the charge conjugate decay mode is implied
unless otherwise stated.
\end{thebibliography}
\end{document}